# The (not so simple!) chain fountain

Rogério Martins[1]
Centro de Matemática e Aplicações
Faculdade de Ciências e Tecnologia, Universidade Nova de Lisboa
2829-516 Caparica, Portugal
E-mail: roma@fct.unl.pt

## Abstract

Given a sufficiently long bead chain in a cup, if we pull the end of the chain over the rim of the cup, the chain tends to continuously 'flow' out of the cup, under gravity, in a common siphon process. Surprisingly enough, under certain conditions, the chain forms a fountain in the air! This became known as the Mould effect, after Steve Mould who discovered this phenomenon and made this experiment famous on YouTube [1], in a video that went viral. The reason for the emergence of this fountain remains unclear. This effect was shown [2] to be due to an anomalous reaction force from the top of the pile of beads, a possible origin for this force was proposed in the same paper. Here, we describe some experiments that give a contribution towards the clarification of the origin of this force, and show that the explanation goes far beyond the one proposed in [2].

**Introduction**

The behaviour of a moving chain could be fairly counterintuitive; this is mainly due to the fact that a chain is a whole connected body, while our intuition is tempted to see it as a set of parts. One of these situations arises in the so-called chain fountain. Given a sufficiently long chain inside a cup that is placed highly enough, if we pull the free end of the chain over the rim of the cup, then the chain continuously pours from the cup to the floor, in a siphoning process. Surprisingly enough, the chain does not simple go around the rim of the cup, a chain fountain forms, see fig. 1. The same phenomena were observed in a continuous rope [3]. This phenomenon was discovered by Steve Mould while searching for a device to explain the behaviour of some polymeric materials [4], later this experiment went viral on YouTube [1].

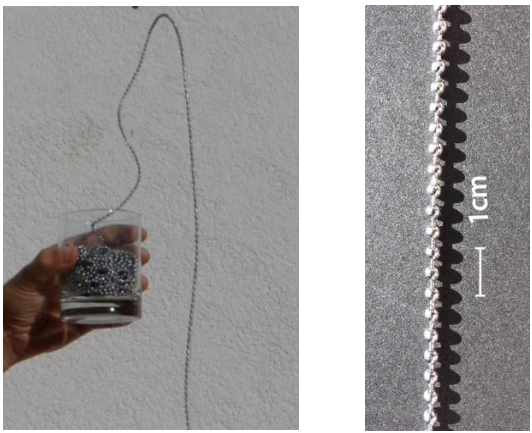

Fig. 1: The chain fountain. All the experiments in this paper were done with a 25 meter (3.2 mm) silver-plated metal bead chain in the image.

[1] This work was partially supported by the Fundação para a Ciência e a Tecnologia (Portuguese Foundation for Science and Technology) through the project UID/MAT/00297/2013 (Centro de Matemática e Aplicações)

The reason for all this mysterious behaviour remains, to some extent, unclear. At first sight we could be tempted to believe that the chain is going upwards because of some inertia of the part of the chain that is going up, to get around the rim, or due to a centrifugal force applied on the beads that are doing the turn. However, as was shown in [2], this is not the case. This leads to the conclusion that the chain is being pushed upwards by a strange reaction force from the top of the pile [2]. In the same paper, there is experimental evidence that the altitude of the fountain is proportional to the altitude of the falling, and the velocity of the chain in the steady state.

On the other hand, there is still no clear understanding of the origin of this reaction force. In the same paper [2], and on the video [5] by the same authors, a possible origin for this force was proposed. Motivated by the fact that the chain has a maximum local curvature, see fig. 2, (a), and the fact that the phenomenon does not happen with a chain that does not have this maximum bending curvature [5], the chain was modelled as a sequence of freely joined rods, see fig. 2, (b). When a rod is pulled from one end, the end attached to the part of the chain that is already moving, the rod tends to rotate around its center of mass, consequently the other end will go down, towards the surface of beads, kicking the pile of beads, see fig. 2, (b). Finally, there is a reaction force to this kick, that pushes the beads upwards. This is, until now, the unique proposed explanation for the origin of the force that causes the chain fountain.

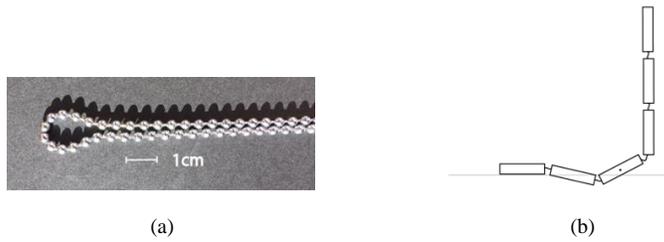

(a)          (b)

Fig. 2: Maximum curvature of the chain and a model suggested in [2].

In this paper, we describe some experiments that show that the proposed cause for the reaction force in [2] gives at most a residual contribution to the chain fountain. On the other hand, some video frames of a moving chain are shown, that give a deeper understanding of the origin of this force.

We believe that the chain fountain phenomenon is somehow related with - at first sight - a different phenomenon: a chain that is pulled by the table where it is falling [6]. Take a length of this kind of chain and suspend it at one end: when the top end is released, and after the bottom end reaches the floor, the free end at the top accelerates faster than a freely falling body. Somehow the furling process occurring at the floor level sucks up the chain. The same phenomenon was studied in [7] where a special chain was tailored to exhibit this behaviour. Again, it was experimentally shown in [6] that the maximum curvature is essential in this phenomenon. Our intuition would say we are observing here the opposite phenomenon to the chain fountain: when the chain is leaving the pile, we have a reaction force upwards, when the chain is entering the pile, we have a force downwards that adds to gravity. In some sense the two phenomena seem to be complementary. We are tempted to think that, if the process could be modeled over time with a differential equation, then the model would be the same for both phenomena: if one is described with time increasing, the other described with time reversed.

The fact that the chain is inside a cup is irrelevant as can be seen in fig. 3. In [8] the particular case of a chain arranged in a table, in one layer, going transversally back and forth along a band, was studied. When one of the ends of this chain is pulled with a great velocity, for example if the chain is falling from a high altitude after it turns over the edge of the table, then the same fountain appears in an arch next to the beads that are initiating their movement. The same phenomenon was observed to happen with a continuous rope [9]. The chain fountain is produced even if the chain is laying more or less randomly inside the tray, in fig.

3 we can see the initial disposition of the chain in the tray. However, as we will explain in the next section, the disposition of the chain is not irrelevant.

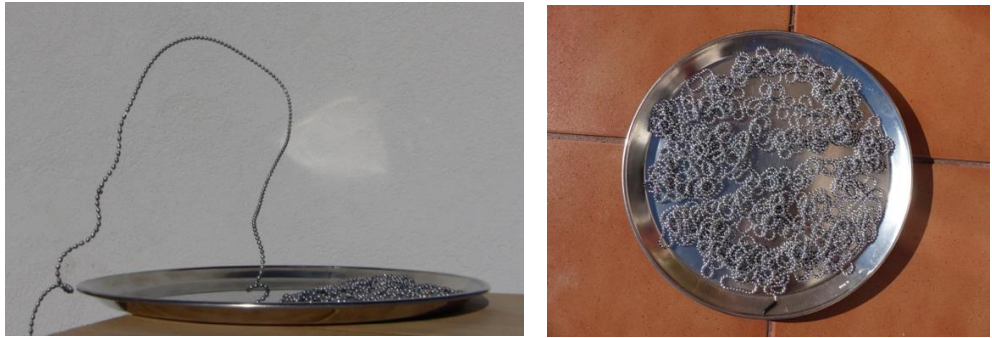

Fig. 3: Chain fountain in a tray and the initial disposition, the rim of the cup is irrelevant.

**The coiled-up pile**

Of course, the chain must be placed inside the cup so that the leading end of the chain is at the top of the pile, so that it may move upwards uninhibited little by little: during the chain fountain process, in each moment, the portion of the chain that starts moving towards the floor must be free on the top of the pile. The pile must not be stirred in order to maintain the correct order of the beads.

On the other hand, usually the chain is cascaded into the cup in a random way and consequently with many twists, see fig. 4 (a), in this situation we have in fact a chain fountain, see fig. 4 (b). However, let us fill the cup in a different way, in order that the chain lays coiled-up inside the cup (essentially avoiding the twists), see fig. 4 (c). In this case, as can be seen in fig. 4 (d), the fountain is considerably smaller.

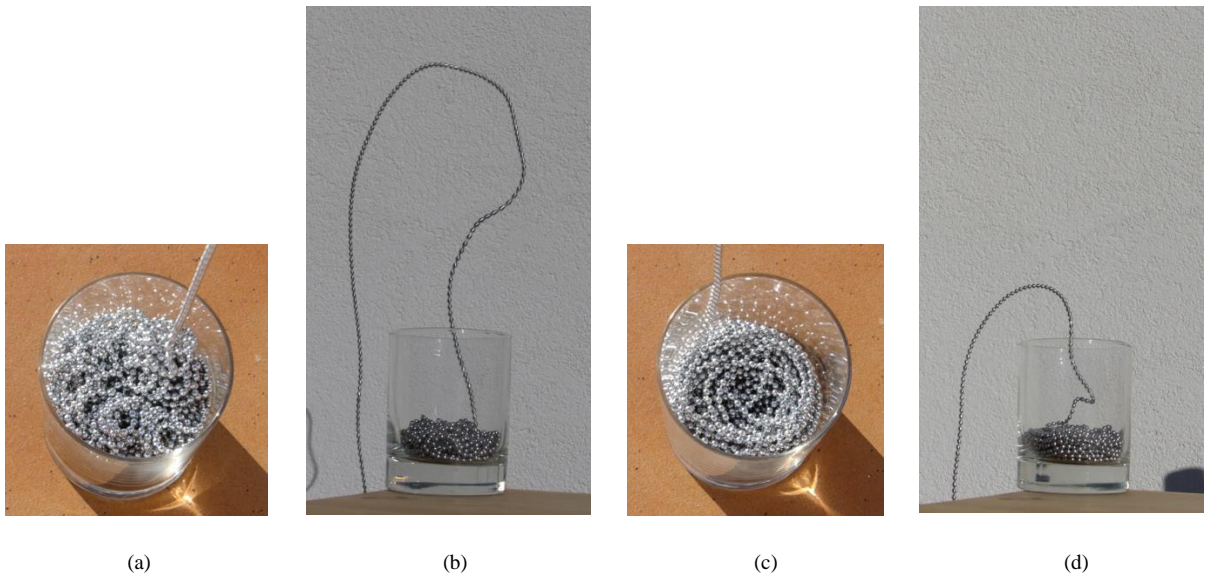

(a)          (b)          (c)          (d)

Fig. 4: In case (a) we see a random filling of the cup, with twists. In case (c) the chain is coiled-up inside the cup. The correspondent process of siphoning can be seen in the other photos: case (c) corresponds to a random filling with twists, case (d) to the coiled-up filling. In both cases the altitude of the falling was 2 meter.

This experiment clearly shows that the conjectured origin of the force creating the chain fountain goes beyond the one given in [2]. Both filling protocols are in line with the assumptions given in [2]. In fact, the coiled-up filling in fig. 4 (c) is even more in accordance with the model given in fig. 2 (b) then the random filling in fig. 4 (a).

**The process seen frame by frame**

If the proposed origin of the force is not the one proposed in [2], what is it then? This section tries to find an answer to this question through some photos taken from video frames of a moving chain. To introduce the method, observe fig. 5 (a). This photo is a frame from a video taken from a bird's eye perspective. The chain was positioned on a table with a contrasting background, the end of the chain that is pointing right is falling onto the floor. Given the gloss of the beads, the trajectory of each bead during the exposure of the frame is visible, given by the lines shown.

Analyzing the subsequent frames we can grasp a lot from the dynamics inside the cup: we are essentially observing a bi-dimensional slice of the chain fountain dynamics inside the cup. The first thing we observe is that the 'bottom' of 'the pile' is being thrust backwards, when the rightmost end is falling onto the floor, see fig 5 (d), to highlight this feature a vertical fixed line was added. Somehow the moving chain in this uncoiling process is pushing 'the pile' backwards. If there was an obstacle behind 'the pile' then there would be a reaction force. This is the force predicted in [2], the force that creates the chain fountain. Another thing we notice is that this force, that pushes 'the pile' backwards, is not constant, it is applied just at some points, some examples of these kicks are highlighted by arrows in fig. 5, (c)(d).

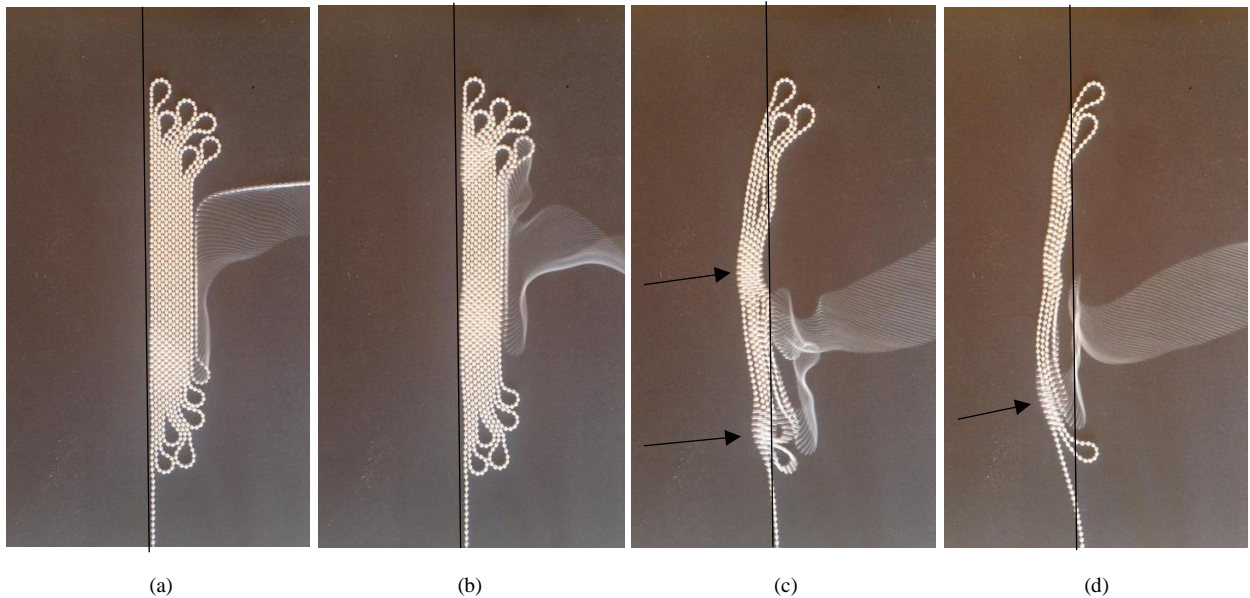

(a) (b) (c) (d)

Fig. 5: Frames taken from a video filmed from a bird's eye perspective. The chain is placed on a table and the rightmost end is falling onto the floor.

To clarify this uncoiling movement, let us start with the simplest non-trivial arrangement we can. The setup is the same but the chain is prepared in an L configuration, see fig. 6 (a). Observing the next frame, fig. 6 (b) we can see that the beads near the corner start to move in a diagonal direction, this is due to the inertia of all the perpendicular row of beads, that was beforehand still.

Another thing we can observe is that there seems to be a stage of acceleration for each bead. In a free fall of a single bead, from zero velocity, there is an acceleration during the fall. However, in a chain fountain, the absolute velocity of the bead that is reaching the floor is precisely the same as the bead that is leaving the cup. At first glance we could be tempted to think that each bead is stationary inside the cup at one moment, and travelling with a positive velocity - the velocity at which the beads are reaching the floor - the next moment. But, of course, this would imply an infinite acceleration at some point in time, something that cannot happen in reality.

The real chain somehow tries to accommodate this abrupt change in velocity with all those strange dynamics that can be observed in fig. 5 (b)(c)(d). In particular, as can be seen in fig 6 (b), there is a preparatory acceleration of the perpendicular row in the very same perpendicular direction.

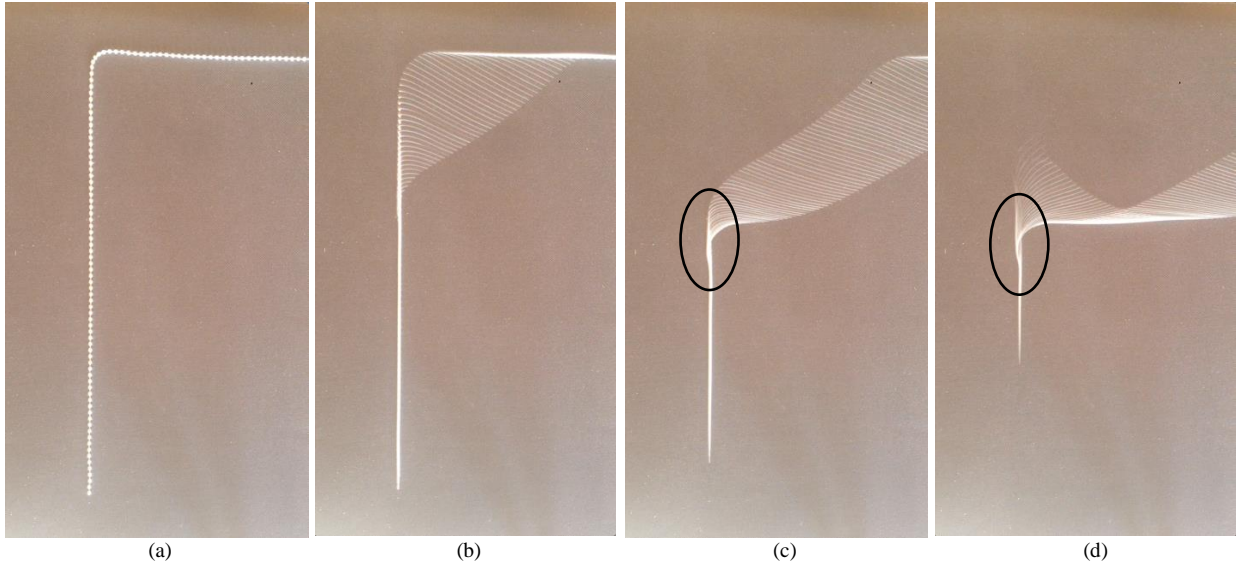

Fig. 6: Frames taken from a video filmed from the bird's eye perspective. The chain is placed on a table and the rightmost end is falling onto the floor.

It is also possible to see, highlighted in fig. 6 (c)(d), a slight backlash before the turn. This is probably due to the bound on the maximum curvature the chain can accommodate, see fig. 2 (a), and is the most similar phenomenon to the kick predicted in [2] we could observe.

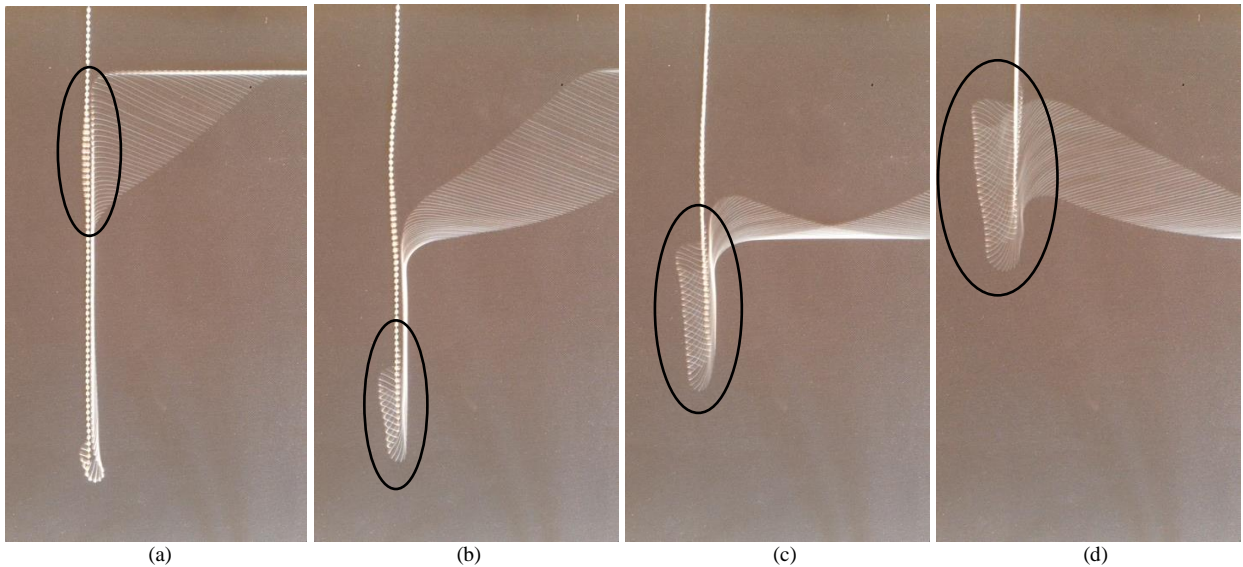

Fig. 7: Frames taken from a video filmed from the bird's eye perspective. The chain is placed on a table and the rightmost end is falling onto the floor.

Let us go one step further and - maintaining the same setup - start with two lines of chain and one twist, see fig. 7 (a). Here we can see a small push on the second row, highlighted in fig. 7 (a), created by the kick we observed in fig. 6 (c)(d). However, in the next frames we see something completely new, there is a huge backlash after the twist, highlighted in fig. 7 (b)(c)(d). The twist creates a whole new dynamic here; the twist somehow creates a kind of free end for the chain, and we have here a sort of whip effect.

If we do the experiment with three lines of chain, see fig. 8 (a). We can observe the push created by this huge backlash in the subsequent video frames. In the beginning, see fig 8 (b), we can see a minor push created by the backlash we see in fig. 6 (c)(d). Later, after the beads in the twist start to move, we can see the push created by the second backlash, highlighted in fig. 8 (c)(d).

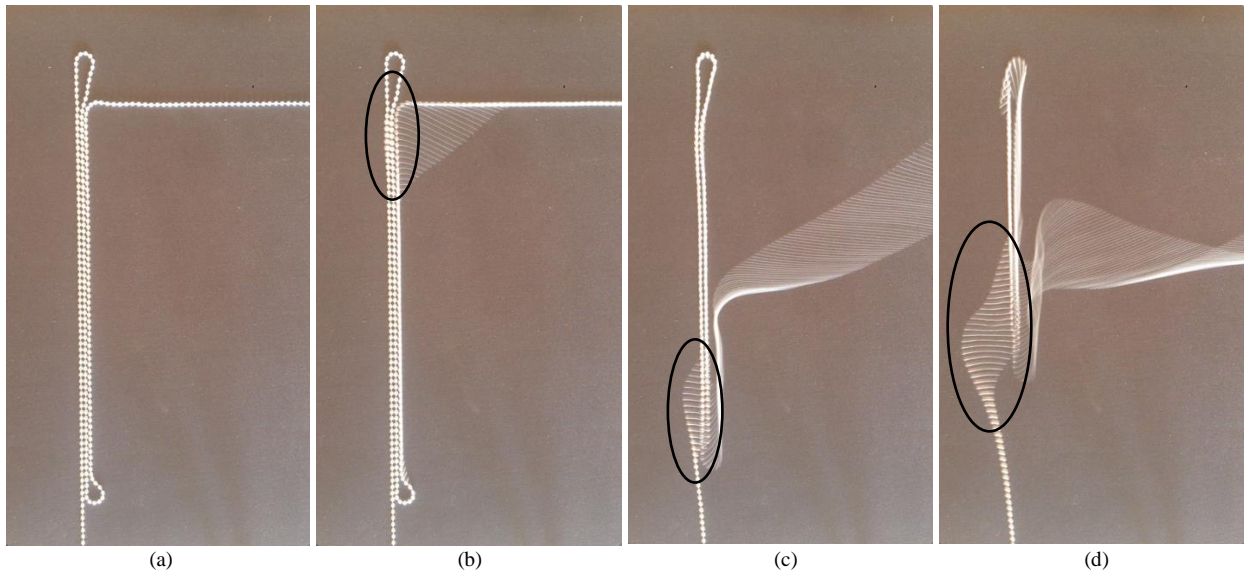

(a)　　　　　　　　　　(b)　　　　　　　　　　(c)　　　　　　　　　　(d)

Fig. 8: Frames taken from a video filmed from the bird's eye perspective. The chain is placed on a table and the rightmost end is falling onto the floor.

We believe that the reaction force that forms the chain fountain is mainly a consequence of this second type of backlash, after the twist. This idea is reinforced by the fact that the chain fountain is bigger when the chain form many twists inside the cup.

**Conclusion**

The observation that the twists inside the cup are determinant for the chain fountain, implies that the proposed model for the reaction force to appear, given in [2], that creates the chain fountain, is too rudimentary to explain the phenomena. As can be seen in the video frames in the next section, the dynamics near the uncoiling zone seems to be much more complex. Apparently, the real push occurs in kicks created after each twist in what resembles a whip effect.

From these images, it is clear now why the coiled-up chain, see fig. 4 (c)(d), forms a smaller chain fountain than the random twisted chain, see fig. 4 (a)(b): if the chain is coiled-up, the process inside the cup is much more like the setting of fig. 6, there is a slight backlash but the push it creates is not strong enough to make a considerable thrust into the air. When the twists are introduced, see fig. 4 (a), we start to have for each twist the strong pushes given by the sort of whip effect we see in fig. 7 (b)(c)(d).

This paper takes the understanding of the origin of this reaction force a bit further, even if the problem is far from understood. More numerical and experimental studies are needed until a complete explanation is reached.

**References:**

[1] Mould S., Self-siphoning beads, (2013). http://stevemould.com/siphoning-beads/

[2] Biggins J. S., Warner M., Understanding the chain fountain, Proc. R. Soc. A, 470 (2014).

[3] Blundell J., Machado H., Fink T., Rope siphon, (2013).
http://www.youtube.com/watch?v=X7CXzjFVUHQ

[4] Mould S., Investigating the "Mould effect", TEDxNewcastle, (2014).
https://www.youtube.com/watch?v=wmFi1xhz9OQ

[5] Biggins J. S., Warner M., Understanding the chain fountain, (2014).
https://www.youtube.com/watch?v=-eEi7fO0_O0&t=472s

[6] Hamm E., Géminard J-C., The weight of a falling chain, revisited, Am. J. Phys. **78**, 828–833, (2010).

[7] Grewal A., Johnson P., Ruina A., A chain that speeds up, rather than slows, due to collisions: how compression can cause tension, Am. J. Phys. **79**, 723–729, (2011).

[8] Hanna J., Santangelo C., Slack dynamics on an unfurling string, Phys. Rev. Lett. **109**, 134301, (2012).

[9] Savage A., Hyneman J., Tablecloth chaos, (2010). http://www.imdb.com/title/tt1747319/